\documentclass[12pt]{article}
\usepackage{amsmath, amssymb, latexsym}
\begin{document}
\bibliographystyle{plain}
\newcommand{\di}{\displaystyle}
\newcommand{\reff}[1]{(\ref{eq:#1})}
\newcommand{\labl}[1]{\label{eq:#1}}
\newcommand{\prob}{\operatorname{prob}}

\centerline{\bf\Large A Simple Solution of the Arrival Time Problem}

\vspace{6mm}

\begin{center}
P. Hrask\'o\\
Department of Theoretical Physics, University of P\'ecs\\
Ifjus\'ag u. 6, 7624 P\'ecs, Hungary
\end{center}

\vspace{4mm}

{\bf Abstract:} Based on the principle that arrival time and position 
are simultaneously measurable quantities a simple formula is
derived for the arrival time probability density 
in nonrelativistic quantum theory.

\vspace{6mm}

The observation of time correlations belongs to the most basic type of
physical experiments. Imagine a beta-radioactive nucleus $_ZX^A$ whose
product nucleus $_{Z+1}Y^A$ decays by alpha-emission. In the most simple
experiment one establishes an alpha-detector at a distance $r_D$ from
the nucleus whose click signalizes the arrival of the alpha-particle. An
electron detector in the immediate vicinity of the nucleus provides the
signal of the {\em moment of preparation} of the nucleus $Y$. The time
$t$ elapsed between the signals of the two detector is called {\em the
arrival time} of the alpha-particle. Notice that this time interval is
typically very much longer than the time $t_D$ which would be required
to cover the distance $r_D$ by the alpha-particle with its mean
velocity. One is inclined to think that the alpha-particle appears only
at the moment $(t - t_D)$ of the decay of $_ZX^A$ but in the situation,
we are considering, no empirical significance can be attributed to the
"moment of decay" itself: It can only be inferred from the value of $t$ provided
the velocity is known. Therefore, the only consistent point of view
is to admit that the particle has been on its way to the point at $r_D$
throughout the whole time interval $t$.

In spite of the fundamental significance of observations of this kind in
quantum theory we do not have any rule to calculate arrival time
probability densities\footnote{A detailed explication of this problem is
found in {\em G. R. Allcock}, The Time of Arrival in Quantum
Mechanics, {\em Ann. Phys.}, {\bf 53}, 253, 286, 311 (1969)}. Assume that the 
alpha-emission can be described by a single-particle potential model and let
us choose the signal of the beta-detector for the zero moment of time.
Then at $t = 0$ the wave-function $\psi ({\bf r},t)$ of the 
alpha-particle will be concentrated within a spherical potential wall around
the point $r = 0$  where the nucleus is found. This wave function is the
only information we have on the alpha-particle. How to deduce from it
the time distribution of the clicks of the alpha-detector?

In order to concentrate on the time behaviour alone without complexities
of the angular distribution we assume isotropic initial conditions which
lead to spherically symmetric wave-function $\psi (r,t)$. Accordingly,
the pointlike detector will be replaced by a thin spherically symmetrical 
shell of mean radius $r_D$. Let us denote the arrival time probability 
density\footnote{No notational distinction will be made between time
and arrival time since the latter will be referred to the zero moment.}
by $\prob (t|r_D,I)$.
The function $\prob (t|r,I)$ is a conditional probability density of the random
variable $t$ which is normalized as\footnote{The integral may diverge
at $t_{min} = 0$. This is a spurious effect due to
the instantaneous spreading of wave-packets in nonrelativistic quantum 
theory and may be avoided if
the considerations are confined to the domain $t > t_{min}$ where
$t_{min}$ is much larger than $r/c$ ($c$ is the speed of light) but
still much smaller than the lifetime.}
\[\int_{t_{min}}^\infty\:dt\;\prob (t|r,I) = 1.\]
The unspecified condition $I$ contains the relevant informations on the protocol 
of the experiment as described above. The coordinate of the detector
which is fixed during the time correlation measurement must be classified also among the
conditions of the experiment and no normalization in it is required.

The problem is how to calculate the
probability density $\prob (t|r,I)$
from the known $\psi (r,t)$ . Standard rules of the quantum
theory connect $\psi$ with the probability density 
\begin{equation}
\prob (r|t,I) = \vert\psi (r,t)\vert^2,\labl{a}
\end{equation}
of the space distribution at a fixed moment of time rather than with $\prob
(t|r,I)$ and presume the normalization condition
\[\int d^3x\:\prob (r|t,I) = \int d^3x\:\vert\psi (r,t)\vert^2 = 1.\]
Reflecting on the radioactive decay (or on the motion of the 
wave-packets), one is often inclined to identify
the arrival time probability density $\prob (t|r,I)$ at a given $r$
with the spatial probability density $\prob (r|t,I)$ (as given is
\reff{a}) at a definite moment
of time but without justification 
this step would be grossly in error. From the elements of probability theory it is
well known that the probabilities $\prob (a|b)$ and $\prob (b|a)$ are
as a rule different from each other. Moreover, in quantum theory 
the time plays the role of the parameter and probabilities are
only defined for dynamical variables at definite moments of time. It is
just the parametric nature of the time which makes arrival time such an 
awkward problem.

In spite of all this, physical properties of the absolute square of the
wave-function $\psi (r,t)$ strongly suggest that
$\prob (t|r,t)$ is indeed proportional to $\vert\psi (r,t)\vert^2$. 
At a fixed $r$ and for times of the
order of the lifetime $\tau$ $\vert\psi (r,t)\vert^2$ decreases in time
approximately as $\exp (-t/\tau)$ while its maximum gets farther and farther
from the origin\footnote{{\em L. Fonda, G. C. Ghirardi and A. Rimini}, 
Decay theory of unstable quantum systems,
{\em Rep. Prog. Phys.}, {\bf 41}, 587 (1978).}.
It is not normalized in $t$ to unity but can be done so. In
other words the prescription
\begin{equation}
\prob (t|r,I) = \frac{1}{N(r)}\vert\psi (r,t)\vert^2\labl{b}
\end{equation}
together with the normalization factor
\begin{equation}
N(r) = \int_{t_{min}}^\infty\:dt\;\vert\psi (r,t)\vert^2\labl{c}
\end{equation}
would, from the observational point of view, provide an excellent
description of the arrival time probability density.

It seems to me that these formulae can be justified from the theoretical
point of view too, accepting the more than plausible principle that {\em
the arrival time and position are simultaneously measurable quantities}.
One might think that this principle contradicts the uncertainty relation
between position and momentum but it does not. From the experiment we
are considering (i.e. from a measurement of the decay law of a
radioactive nucleus) no information can be inferred on the momentum of
the emitted particle. A more general argument is that, using
time of flight spectrometers, we actually measure the position in two
subsequent moments of time rather than momentum itself. Substantial
additional knowledge (the absence of a force field along the path)
is required {\em to infer} the value of the momentum prior to the detector
response. 

What the above principle does exclude is that arrival time is an
operator in the Hilbert-space which does not commute with the coordinate
operators. But this mathematical property of the observables under study
could only be maintained if an
experiment designed to measure precisely either of them excluded the
possibility to measure the other. No such principal incompatibility
seems to exist for the simultaneous observation of the arrival time and
position.  

If arrival time and position are indeed simultaneously measurable
entities then their joint probability density $\prob (r,t|I)$ is
a sensible quantity whose existence permits us to introduce the
conditional probability densities $\prob (t|r,I)$ and $\prob (r|t,I)$ as
\begin{align}
\prob (t|r,I) & = \frac{\prob (r,t|I)}{\prob (r|I)},\labl{d}\\[2mm]
\prob (r|t,I) & = \frac{\prob (r,t|I)}{\prob (t|I)}.\labl{e}
\end{align}
On the first line $\prob (r|I)$ is the probability density to find the
alpha-particle (under the conditions $I$ of the experiment) in $r$ at
some moment of time:
\begin{equation}
\prob (r|I) = \int_{t_{min}}^\infty\:dt\;\prob (r,t|I) =
\int_{t_{min}}^\infty\:dt\;\prob (r|t,I)\cdot\prob (t|I).\labl{f}
\end{equation}

The function $\prob (t|I)$ on the second line is equal to the
probability density to find the alpha-particle at the moment $t$
somewhere in space. The only natural possibility for this probability is that it does not
depend on time since the particle exists no more in one moment of time
than in another\footnote{For a close analogy
remember that in the elementary theory of the radioactive decay
the decay probability $\lambda\cdot dt$ in the interval $(t,t+dt)$ 
is independent of $t$.}.Since a uniform density is not normalizable in the
semiinfinite interval $t_{min} < t < \infty$ we are compelled to confine
ourselves to an arbitrarily large but finite time interval $t_{min} < t <
T$. Then $\prob (t|I)\approx 1/T$, but the arbitrary parameter
$T$ drops out of the final formula.

Now, eliminating $\prob (r,t|I)$ from \reff{d} and \reff{e} we obtain
the Bayes-like formula
\[\prob (t|r,I) = \frac{\prob (r|t,I)\cdot\prob (t|I)}{\prob (r|I)}.\]
Inserting \reff{a}, \reff{f} and $\prob (t|I) = 1/T$ into this equation, we arrive
at the desired result \reff{b}, \reff{c} which seems to be the most natural
solution of the arrival time problem. 

Equations \reff{b} and \reff{c} are applicable also in the case when a
high potential barrier is present between the decaying nucleus and the
detector so that the particle can reach the detector only through
tunneling. Properties of the tunneling time\footnote{For a review, see
{\em R. Landauer and Th. Martin }, Barrier interaction time in tunneling,
{\em Rev. Mod. Phys.}, {\bf 66},  217  (1994).} can then be studied by
calculating $\prob (t|r,I)$ both in the presence of the barrier and
without it.

The above considerations have to be implemented by the detector
efficiency $\epsilon$. Following common practice, the theoretical
probability distribution \reff{b} must be multiplied by $\epsilon$ which
takes values in the interval $(0,1)$: In $N$ trials the detector
responds in $\epsilon N$ cases. If it is desirable to protect the 
wave-function from distortions due to the presence of the detector the
efficiency must be close to zero.

In justifying \reff{b} I have been guided by the conviction that this
formula seems capable of reproducing the properties of the arrival time
known from experience. Its "simplicity" is the natural consequence of
this effort but I hope a
sufficiently solid foundation has been given to it. On the other hand, my proposal
is connected also to the state reduction hypotheses which is probably the only
problem within quantum mechanics which has remained controversial since
the time of its birth. The recipe \reff{b} does not circumvent this
problem: When the detector clicks the wave-function $\psi (r,t)$
collapses into the domain of the detector and this process is outside
the scope of the Schr\"odinger-equation. Yet this recipe contains an
essentially new element since the moment of the collapse is now "chosen
by the system itself" rather than by the (hypothetical) intervention of
the observer.

The author is deeply indebted to A. Shimony for his criticism of
an early version of the paper. 

\end{document}